\documentclass[english,twocolumn,showpacs,prb]{revtex4}

\usepackage[T1]{fontenc}
\usepackage[latin9]{inputenc}
\usepackage{graphicx}
\usepackage{latexsym}
\usepackage{stmaryrd}

\usepackage{color}
\usepackage{babel}

\begin{document}

\title{A controllable valley polarization in Graphene}
\author{Yang Liu$^{1}$}
\author{Juntao Song$^{1}$}
%\author{Juntao Song$^{1,\ast}$}
\email{jtsong@hebtu.edu.cn}
\author{Yuxian Li$^{1}$}
\author{Ying Liu$^{1}$}
\author{Qing-feng Sun$^{2}$}
\affiliation{$^1$Department of Physics and Hebei Advanced Thin Film  Laboratory, Hebei Normal University, Hebei 050024, China\\
$^2$Institute of Physics, Chinese Academy of Sciences, Beijing 100080, China \\
}

\date{\today}

\begin{abstract}
The electron transport of different conical valleys is investigated in graphene with extended line-defects. Intriguingly, the electron with a definite incident angle can be completely modulated into one conical valley by a resonator which consists of several paralleling line-defects. The related incident angle can be controlled easily by tuning the parameters of the resonator. Therefore, a controllable $100\%$ valley polarization, as well as the detection of the valley polarization, can be realized conveniently by tuning the number of line-defects and the distance between two nearest neighbouring line-defects. This fascinating finding opens a way to realize the valley polarization by line-defects. With the advancement of experimental technologies, this resonator is promising to be realized and thus plays a key role in graphene valleytronics.
\end{abstract}

\maketitle
\section{Introduction}
Since the successful fabricating of graphene in 2004,~\cite{Novoselov0} its excellent transport properties,~\cite{Novoselov,Yzhang,Geim1,Geim2} such as the exceptional electron and thermal transport properties at room temperature, have attracted a lot of research interest to apply it in making future electronic devices.~\cite{Avouris,Neto0,Bolotin,Balandin,Abergel,Schwierz} Many ideas and attempts have been going on. One of the most fundamental and revolutionary attempts is to control the valley degree of freedom to realize the signal's storage and transport.\cite{Beenakker,KChang,Zeng,YJiang,Zhai} Supposed an achievement of this idea in future, graphene will probably replace all semiconductors in conventional electronics and bring a new era in many fields. Before the coming of the new era, there is a long way to improve the control over transport in graphene.

Inspired by the crucial role of the doping in semiconductor, many researchers have studied the effects related to disorder in graphene,\cite{Neto0,aSun,Jiang,YanyangZhang} which consists of adsorbed atoms (or molecules), charged impurities, vacancies, or other topological defects. It is anticipated to modify the transport property of pristine graphene to fulfill the requirement. Recently, a peculiar topological line-defect in graphene was reported experimentally by Lahiri et al.~\cite{Lahiri} This topological line-defect is created by alternating the Stone-Thrower-Wales defect and divacancies, leading to a pattern of repeating paired pentagons and octagons, as shown in Fig. 1. In particular, it is found in this experiment that this line-defect has metallic characteristics. Great attention has been attracted to the line-defect and further investigations manifest a variety of its promising applications.\cite{Song1,Alexandre,Bahamon,Okada,Kou,Gunlycke}

Although a lot of theoretic proposals are presented in graphene to realize the valley filter before, all of them are not realized till now due to the extreme difficulty in the experimental implementation. After the discover of line-defect, Gunlycke and White~\cite{Gunlycke} reported that a valley polarization near $100\%$ can be achieved by scattering off a line-defect.\cite{Gunlycke} This original idea subsequently arouse physics community's intensive interests.\cite{Zheng1,Zheng2,Gunlycke2} As a new discovery, the extended line-defect is very promising to realize the valley polarization in graphene. However, because the $100\%$ valley polarization only maintains at a large incident angle in these reported results.\cite{Gunlycke,Zheng1,Zheng2} A serious problem will be encountered in experiment that the electron must always follow the direction of the line-defect to maintain a high valley polarization. This transport characteristic is almost fatal to utilize efficiently the high valley polarization. The purpose of this paper is to address a feasible approach to realize a $100\%$ valley polarization at a small incident angle by using a resonator, which consists of several paralleling line-defects.

Given a high valley polarization by scattering off one line-defect, a question is putted forward naturally: What will happen in the presence of two or more line-defects? Motivated by finding an efficient realization of the high valley polarization, we therefore study the electron transmission coefficient in the presence of several line-defects. Intriguingly, it is found that an obvious change happens to the transmission coefficients with varying the distance between two line-defects. In particular, for the electron with a definite momentum and incident angle, the transmission coefficients can reach to one at a definite distance between two line-defects. Simple numerical analysis argues that the large transmission coefficients should be attributed to the standing wave structure due to the electron's quantum interference, which is also named as Fabry-P$\'{e}$rot interference in optics. Based on a further numerical simulation, an efficient device to realize a $100\%$ valley polarization is further proposed for six line-defect configuration at the end. This fascinating finding opens a way to realize the valley polarization by line-defects.

The rest of this paper is organized as follows: In Sec.~\ref{sec:models}, we introduce the Hamiltonian in the tight-binding model and present the formula of the transmission coefficient. The numerical results are discussed in Sec.~\ref{sec:discussions}. Finally, a brief summary is given in Sec.~\ref{sec:conclusions}.

\section{Theoretical Models}
\label{sec:models}
In the tight-binding approximation, a single layer of graphene with a line-defect, sketched in Fig. 1(a), can be described by the following Hamiltonian:
\begin{eqnarray}\label{M1}
H=t\sum_{\langle\mathbf{i},\mathbf{j}\rangle}c^+_\mathbf{i}c_\mathbf{j}+\tau_2\sum_{\langle\alpha,\beta\rangle}c^+_{i_y,\alpha}c_{i_y,\beta}
+\tau_1\sum_{\langle\mathbf{i},\alpha\rangle}c^+_\mathbf{i}c_{i_y,\alpha}+h.c.,
\end{eqnarray}
where $c_\mathbf{i}$ and $c_{i_y,\alpha/\beta}$ represent the electron
annihilation operators on the graphene site $\mathbf{i}$ and line-defect sites, respectively, and ${\langle\cdot\rangle}$ refers to the nearest neighboring sites. Supposed a variation less than $5\%$ in the hopping terms $\tau_1$ and $\tau_2$,\cite{Song1,Bahamon,Okada,Kou} here we adopt $\tau_1 \approx\tau_2 \approx t=-1$.

\begin{figure}
\center
  \includegraphics[width=0.94\columnwidth]{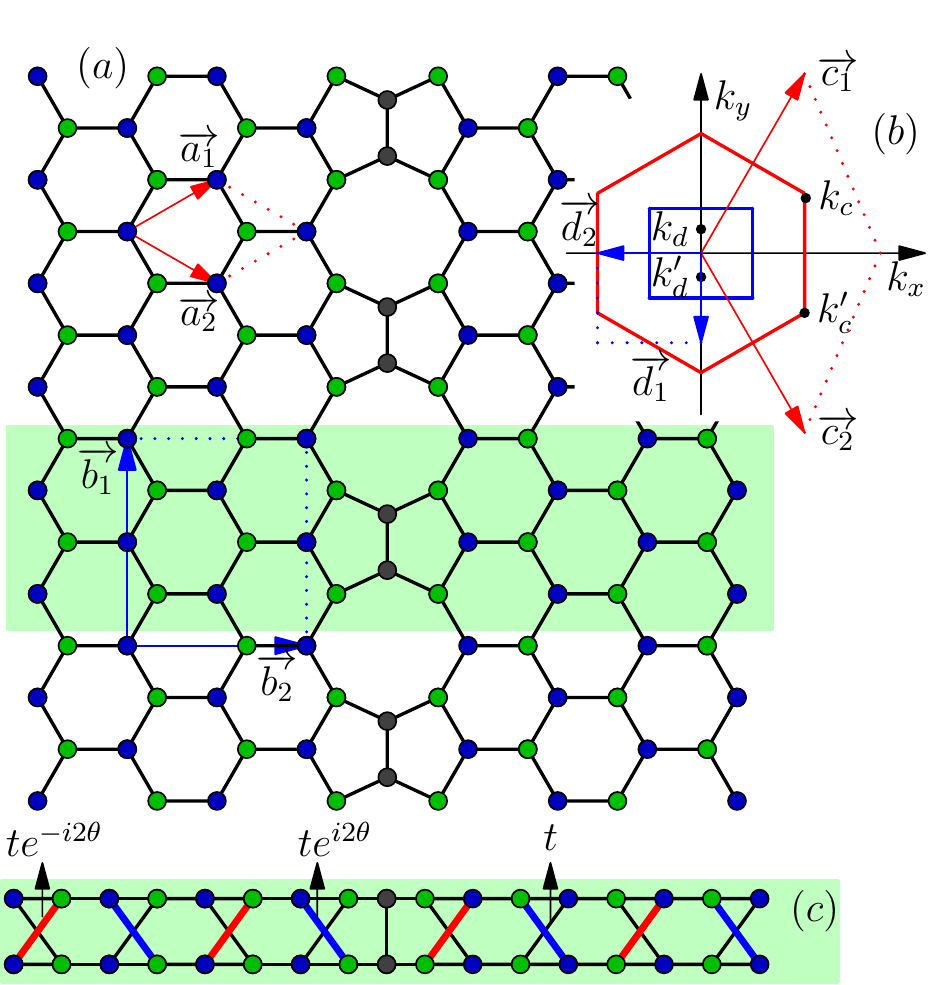}
\caption{ (Color online) (a) The schematic diagram of the graphene sheet with a line-defect
along y axis. (b) The Brillouin zones of Graphene, which correspond the single cell and the supercell shown in (a). (c) The simplified lattice model of the infinite graphene with a line-defect. Here, $\theta=k_ya$ is from the Fourier transformation along y axis. }
\end{figure}

Because of the existence of line-defect, the translation symmetry along $x$ axis breaks down for a graphene system, as sketched in Fig. 1(a). However, the translation symmetry along $y$ axis still retains and therefore $k_y$ is a good quantum number, namely a conserved quantity. For the sake of the convenience in computing, we do the Fourier transform only along $y$ axis as follows:
\begin{eqnarray}\label{M2}
&&c^+_\mathbf{i}=\sum_{k_y}c^+_{k_y,i_x}e^{-i2k_yi_ya},c_\mathbf{i}=\sum_{k_y}c_{k_y,i_x}e^{i2k_yi_ya},\nonumber\\
&&c^+_{i_y,\alpha}=\sum_{k_y}c^+_{k_y,\alpha}e^{-i2k_yi_ya},c_{i_y,\alpha}=\sum_{k_y}c_{k_y,\alpha}e^{i2k_yi_ya},
\end{eqnarray}
$a$ represents the lattice constant in pristine graphene. After the Fourier transformation, the Hamiltonian in Eq. (\ref{M1}) is decoupled into $H=\sum_{k_y}H_{k_y}$. Note that here $H_{k_y}$ can be effectively represented by a quasi-one-dimensional lattice model, as shown in Fig. 1(c). Readily, we can describe the Hamiltonian $H_{k_y}$ as follows:
\begin{eqnarray}\label{M3}
&&H_{k_y} =-\sum_{i}\varphi _{i,1}^{\dagger }\hat{T}_1\varphi _{i,2}
-\sum_{i}\varphi _{i,2}^{\dagger }\hat{T}_2\varphi _{i,3}
-\sum_{i}\varphi _{i,3}^{\dagger }\hat{T}_1^\dagger
\varphi _{i,4}\nonumber\\
&&-\sum_{i\neq -1}\varphi _{i,4}^{\dagger }\hat{T}^\dagger_2 \varphi _{i+\hat{x},1}
-\varphi _{-1,4}^{\dagger }\hat{T}_2 \varphi _{0}\nonumber
-\varphi _{0}^{\dagger }\hat{T}_2 \varphi _{1,1}\nonumber\\
&&-\varphi _{0}^{\dagger }\hat{T}_3\varphi _{0}+h.c.,\\
&&\hat{T}_1=\left(
\begin{array}{cc}
1 & 1 \\
e^{-i2\theta} & 1
\end{array}
\right),
\hat{T}_2=\left(
\begin{array}{cc}
1 & 0 \\
0 & 1 \\
\end{array}
\right),
\hat{T}_3=\left(
\begin{array}{cc}
0 & 1 \\
1 & 0 \\
\end{array}
\right)\nonumber
\end{eqnarray}
where $\theta=k_y a$, $\varphi^\dagger_{i,\alpha}=[c^\dagger_{k_y,i,\alpha,A},c^\dagger_{k_y,i,\alpha,B}]$, and $\hat{x}$ represents the unit length between the neighboring supercells at the graphene part. Here, $i$ represent the position of a supercell, $\alpha$ takes the integer number from 1 to 4 to denote the different columns in a supercell, and $A/B$ in $c^\dagger_{k_y,i,\alpha,A/B}$ corresponds to the up/down site in the same column in Fig. 1(c).

\begin{figure}
\center
  \includegraphics[width=1.0\columnwidth]{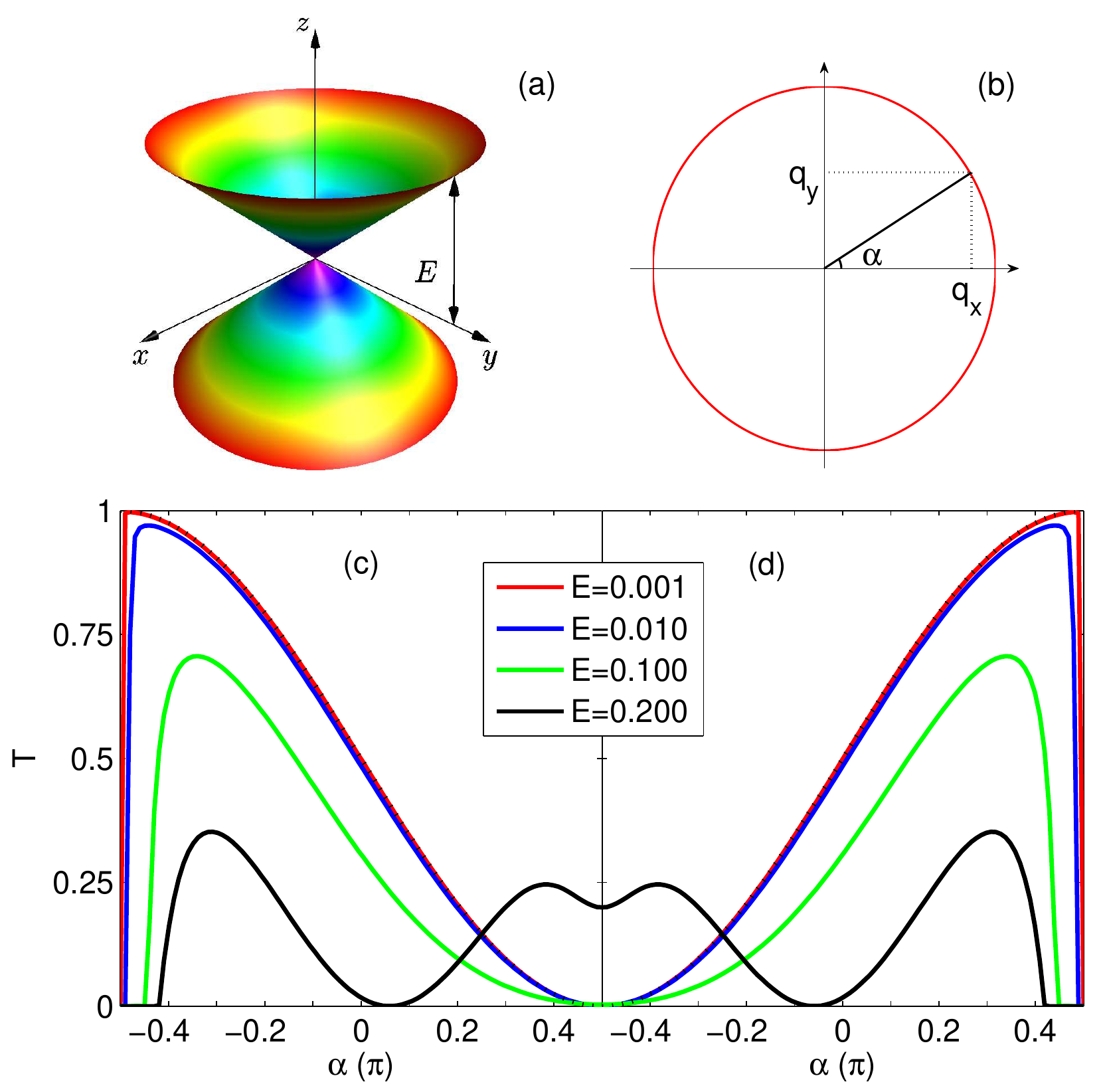}
\caption{ (Color online) (a) The schematic diagram of $K$ or $K'$ Dirac cone . (b) The relationship between the electron's incident angle and momentums. (c) and (d) manifest the transmission coefficient in $K$ and $K'$ valleys. In this case, there is only one line-defect. Here, the dotted line represents the approximate result in the low energy limit: $(1\pm\sin\alpha)/2$}
\end{figure}

According to the Landauer-B\"{u}ttiker formula and the Hamiltonian in Eq. (\ref{M3}), the transmission coefficient of the line-defect system at zero temperature and low bias voltage can be represented as:~\cite{Song2,Pareek,SDatta}
\begin{eqnarray}\label{M4}
T_{k_y} (E)=Tr{[Re(\Gamma_LG^r\Gamma_RG^a)]},
\end{eqnarray}
where $G^{r/a}$ is the retarded/advanced Green's function related to the line-defect Hamiltonian and $\Gamma_{L/R}=i(\Sigma^r_{L/R}-\Sigma^a_{L/R})$ with the retarded/advanced self-energy $\Sigma^{r/a}_{L/R}$. Note that here the left/right lead is represented by a semi-infinite quasi-one-dimensional graphene lattice sketched in Fig. 1(c) and the sample Hamiltonian refers to the line-defect Hamiltonian, namely $\hat{T}_3$.

\begin{figure}
\center
\includegraphics[width=1.0\columnwidth]{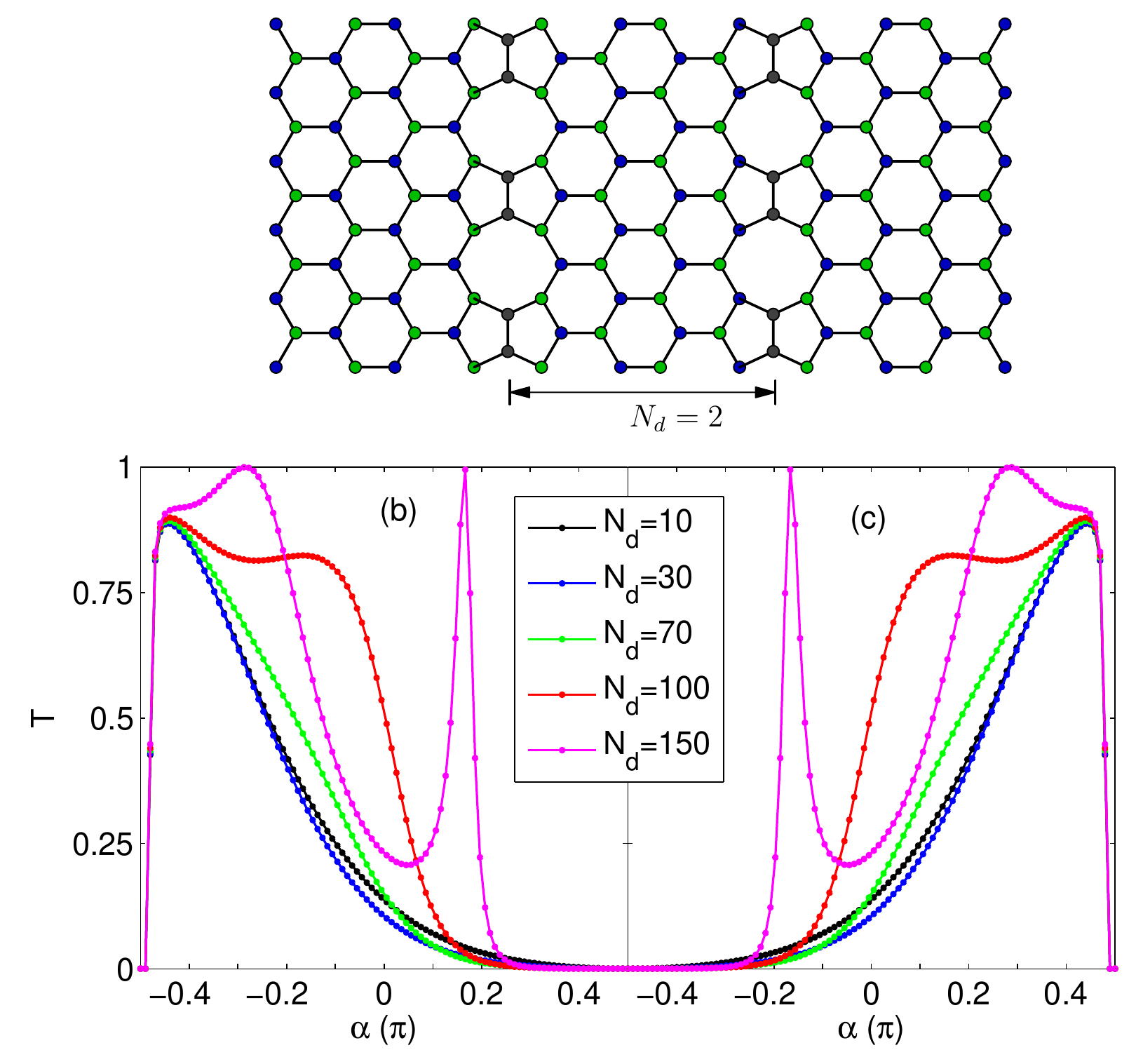}
\caption{ (Color online) (a) The schematic diagram of the graphene sheet with two line-defects. The transmission coefficients in $K$ (b) and $K'$ (c) are mapped as a function of the incident angle $\alpha$ in the presence of two line-defects. Here, electron's energy is set as $E=0.01$. }
\end{figure}
It is worth noting that we can readily compute the transmission coefficients of the electron with a fixed incident angle $\alpha$ in two Dirac points, $K$ and $K'$, by using the above formula. As shown in Fig. 1(b), two Dirac points $K$ and $K'$ which are located at $[0,\pm4\pi/3a]$ for a pristine graphene are now cast at $k_d$ and $k'_d$ for a graphene with a line-defect, $[0,\pm\pi/3a]$. Therefore, an
electron in two valleys satisfies the following relations: $k_x=q_x$ and $k_y=\pm\pi/3a+q_y$, where $q_x/q_y$ represents electron's group velocity along the $x/y$ direction. Combined with the Dirac electron's linear dispersion relation $E=\frac{\sqrt{3}aq}{2}=\frac{\sqrt{3}a\sqrt{q_x^2+q_y^2}}{2}$, the transmission coefficients can be readily mapped as a function of the electron incident angle $\alpha$,\cite{Zheng2} where $\alpha=\arctan(q_y/q_x)$. For simplicity in discussing, we will never distinguish the difference of the Dirac points in two Brillouin zones in the following and uniformly adopt $K$ and $K'$ to represent the two nonequivalent Dirac points.

Besides, some previous papers reported that in the vicinity of the Dirac point, the transmission coefficient of low energy electron will obey the law: $T_{K/K'}=(1\pm\sin\alpha)/2$. In order to make this work essentially self-contained, we also confirm this conclusion and present the related derivations in Appendix.

\section{Numerical Results and Discussions}
\label{sec:discussions}
\subsection{One line-defect}
Even though the transmission coefficient in the vicinity of the Dirac point was given by an approximated method in many references,\cite{Gunlycke,Zheng2} to the best of our knowledge, the exact numerical calculation without any analytical approximations is not unambiguously shown before. Given the realization of the valley polarization using the line-defect in future, what the transmission coefficient behaviors at a relatively large energy is a fundamental question.

In Fig. 2, the transmission coefficient is mapped as a function of the electron's incident angle for different energies. It can be observed the line of $E=0.001$ (red) overlaps exactly with the dotted line, $(1\pm\sin\alpha)/2$, which represents the transmission coefficient of low energy electron in the vicinity of the Dirac point by an approximate method and is also given in Appendix. One subtle difference between the numerical and approximate results is zero transmission probability nearby $\alpha=\pi/2$ for exact numerical result. However, this physical observation is not manifested in the approximate result (dotted line). A severe deviation from the dotted line is clearly seen for the lines with $E\geq0.010$, which indicates that a excellent valley transmission can only be maintained at not very large energy.

\begin{figure}
\center
  \includegraphics[width=1.0\columnwidth]{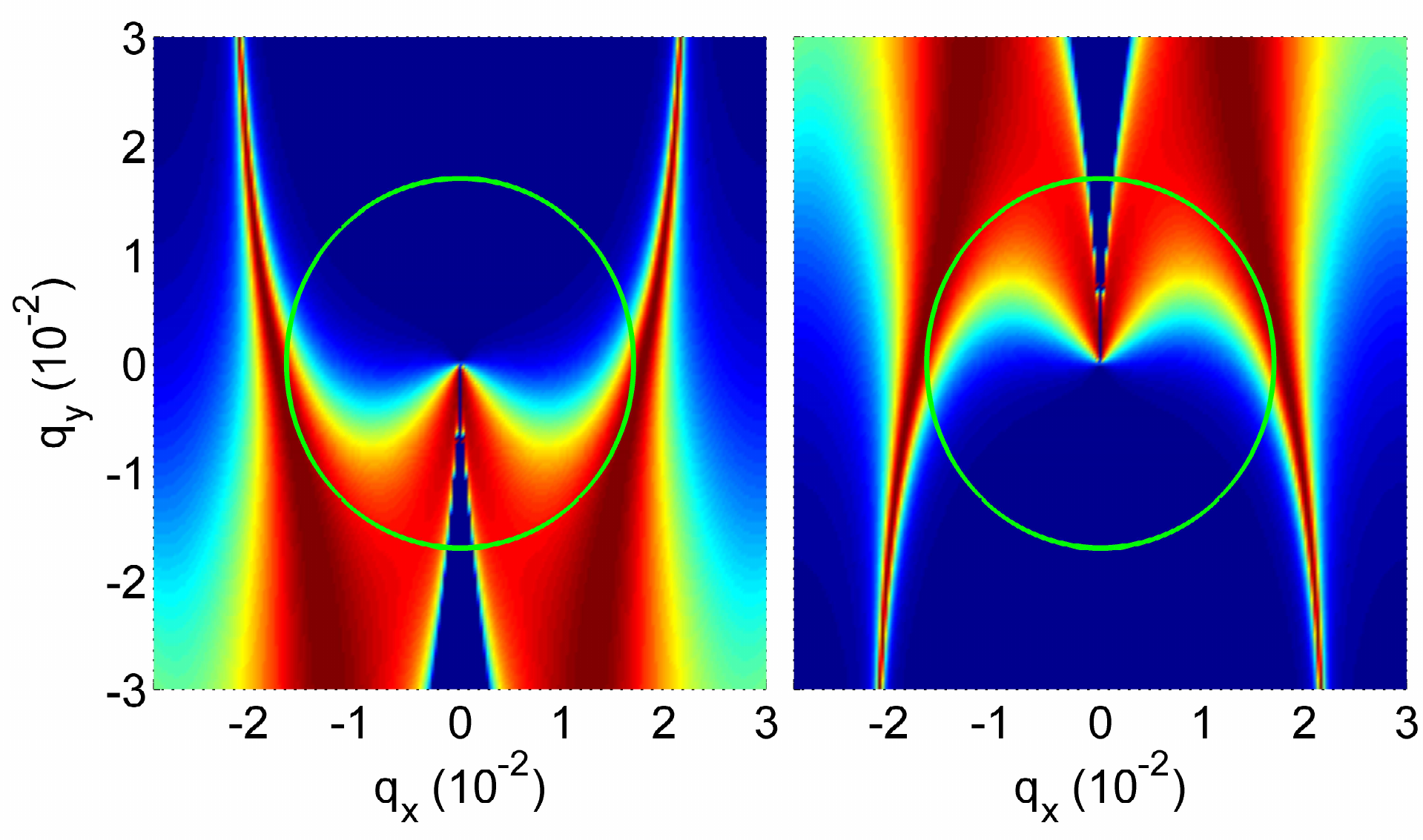}
\caption{ (Color online) Contour plots of the transmission coefficients in $K$ (left) and $K'$ (right) valleys. In this case, two line-defects are considered and the fixed distance between them is $N_d=74$.}
\end{figure}

\subsection{Two line-defects}
The search of the valley polarization material or device attracts physics community greatly. Although a proposal was given recently by scattering off a line-defect,\cite{Gunlycke} it looks much difficult for graphene with only one line-defect to make electron leave away from the line-defect and finally reach the key device, and meanwhile maintain at a high valley polarization in this process. Inspired by the valley polarization in the line-defect graphene, we expect this problem can be resolved by scattering off two or several line-defects.

In Fig. 3, the transmission behavior is mapped as a function of the incident angle in the presence of two line-defects. Fig. 3(a) shows a lattice model of graphene with two line-defects and meanwhile presents a schematic interpretation of $N_d=2$ which represents the distance between two nearest neighbouring line-defects. Using the formula (4), the transmission coefficients in the $K$ and $K'$ valleys are mapped as a function of the incident angle in Fig 3(b) and 3(c). A mirror symmetry with respect to $\alpha=0$ can be seen clearly between them. Comparing to the case of one line-defect, there is almost no distinct change for the lines of $N_d=10/30$ both in Fig. 3(b) and 3(c). However, the transmission coefficient starts varying for $N_d\ge 70$ and in particular some transmission peaks can be observed obviously at small incident angles for the lines of $N_d=100$ and $150$. This exciting finding makes us fascinated to explore what happens to the two-line-defect graphene and why there is a high transmission probability at some incident angles.
\begin{figure}
\center
  \includegraphics[width=0.8\columnwidth]{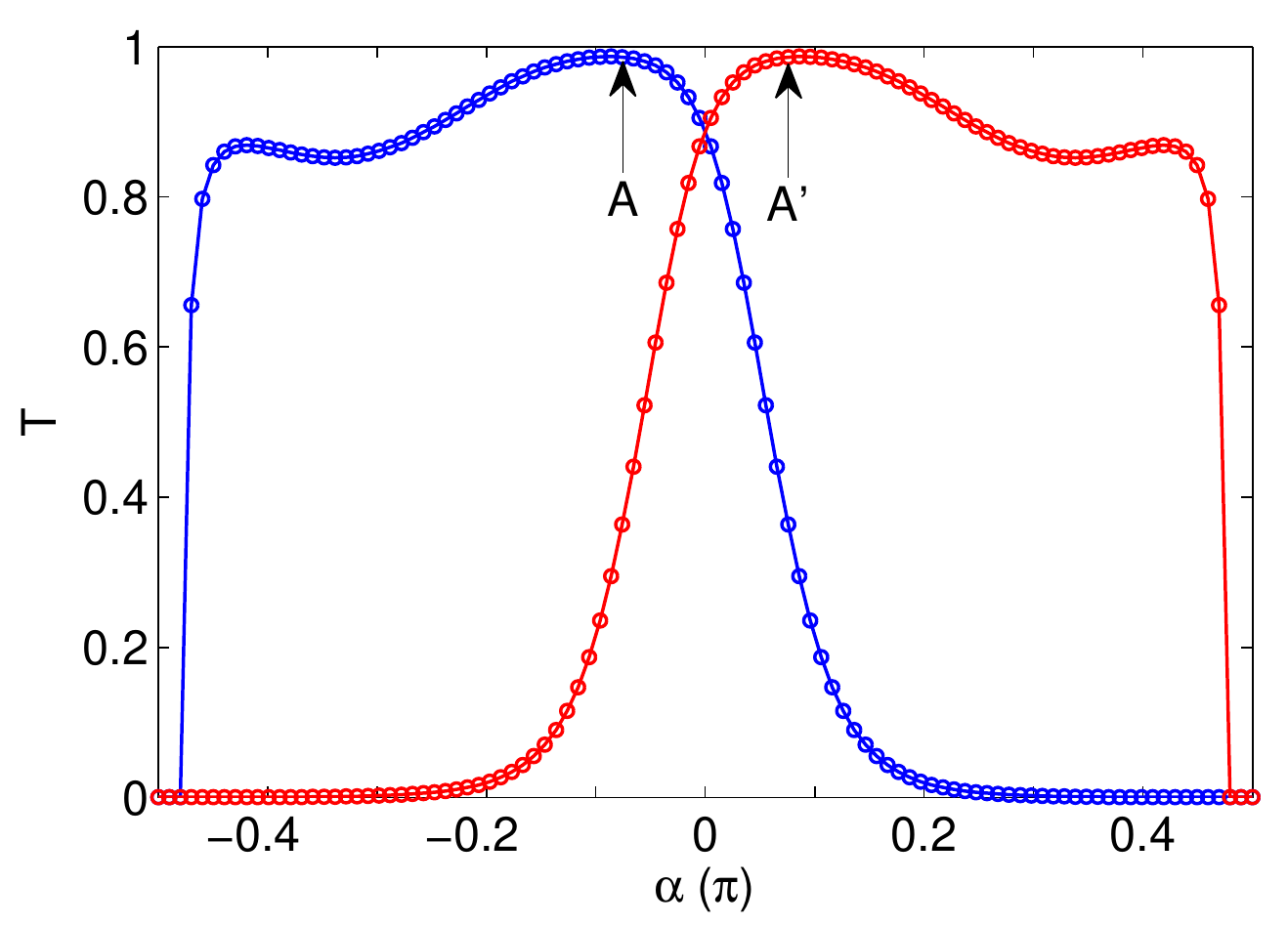}
\caption{ (Color online) The transmission coefficient with a fixed momentum $q=0.017/a$ which is exactly the green circle path in Fig. 4. Two lines (blue and red) correspond to that of $K$ and $K'$ valleys, respectively.  In this case, two line-defects are considered and the fixed distance between them is $N_d=74$.}
\end{figure}

In order to demonstrate a whole physical picture, we map the transmission coefficient as the function of $T(q_x,q_y)$ in Fig. 4. The left and right maps, which correspond to the $K$ and $K'$ valleys, respectively, manifest a clear symmetry. Intriguingly, two large transmission regions (dark red) can be observed both in the $K$ and $K'$ valleys. Note that the positive (negative) $q_x$ represents an incident electron along positive x direction (negative x direction). These peaks remind us some quantum interferences occur in the presence of two line-defects, e.g. electron's resonant tunneling phenomena.

To confirm the origin of the high transmission above, we plot two lines selectively in Fig. 5. These two lines which belong to $K$ and $K'$ valleys respectively correspond to half circle of the green route in Fig. 4, representing $q=\sqrt{q^2_x+q^2_y}=0.017/a$. Obviously, the transmission peaks at $A$ and $A'$ are from the intersection of the green circle route and the high transmission region (dark red). When substituting $q=0.017/a$ and $\alpha_{A/{A'}}\approx\pm0.05\pi$ into the formula: $\lambda=1/E=2/\sqrt{3}qa$ and $l=N_d/\cos{\alpha}$, we find two quantities are equivalent within the physical error range. This is to say, the $100\%$ transmission does originate from the resonant tunneling, which is often observed in the system of quantum dot.\cite{Song3}

\begin{figure}
\center
\includegraphics[width=1.0\columnwidth]{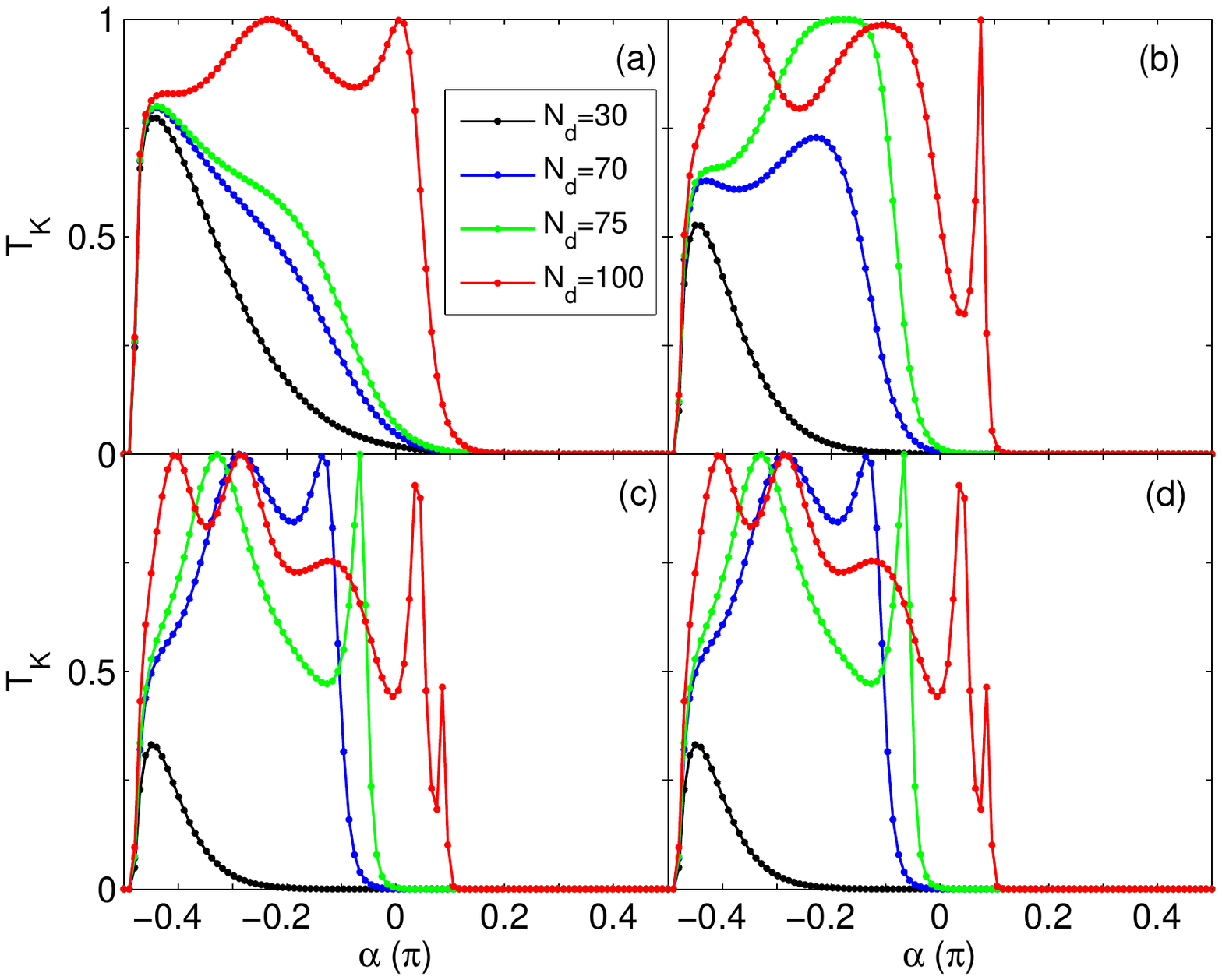}
\caption{ (Color online) Transmission coefficient in only $K$ valley is given in the presence of several paralleling line-defects: (a) $m=3$, (b) $m=5$, (c) $m=7$, and (d) $m=10$. Note that here $m$ represents the number of paralleling line-defects. In each case, four lines with differen $N_d$, the distance of two nearest neighbouring line-defects, are manifested. Here, electron's energy is set as $E=0.01$}
\end{figure}

According to the quantum theory of wave-particle duality, the low energy electron in the vicinity of the Dirac point is inclined to exhibit undulation characteristics. When scattering off two line-defects, a standing wave state between two line-defects helps electron resonantly tunnel through the barriers of the line-defects. Consequently, we can observe this perfect transmission phenomenon in above figures. Using the valley polarization formula $\mathcal{P}=\frac{T_{k}-T_{k'}}{T_{k}+T_{k'}}$, the $100\%$ valley polarization can not be realized at very small incident angle in the presence of two line-defects, e.g. at the positions of the $A$ and $A'$ peaks. Nevertheless, an exciting thing is that $100\%$ valley polarization can almost be reached when  $\alpha$ takes the value of $[0.2\pi,0.3\pi]$, a relatively small incident angle. At least, the results argue that the high valley transmission at a small incident angle is feasible by scattering off two line-defects.

\begin{figure}
\center
  \includegraphics[width=1.0\columnwidth]{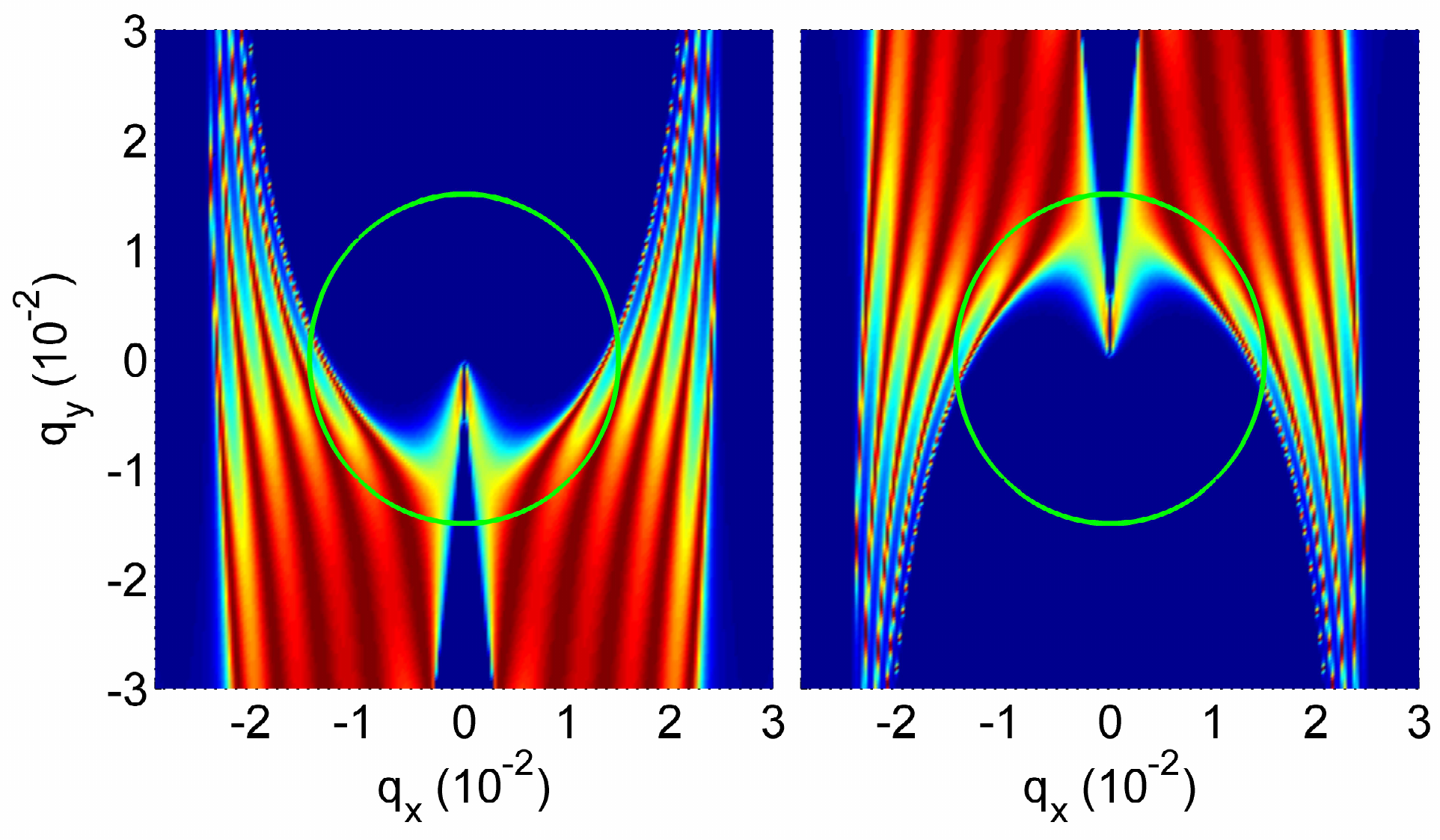}
\caption{ (Color online) Contour plots of the transmission coefficients in $K$ (left) and $K'$ (right) valleys. In this case, six line-defects are considered and the fixed distance between arbitrary two neighboring line-defects is $N_d=70$.}
\end{figure}

\subsection{Many line-defects}
Inspired by the above results, we expect an excellent case could occur to the graphene with several line-defects. In Fig. 6, the transmission coefficient only in $K$ valley is shown in the presence of several paralleling line-defects: (a) $m=3$, (b) $m=5$, (c) $m=7$, and (d) $m=10$, where $m$ represents the number of paralleling line-defects. In Fig. 6(a) with $m=3$, almost no obvious change can be observed with comparison to the case of $m=2$ in Fig. 3. Nevertheless, one extra transmission peak still appears in the line of $N_d=100$ (red). Without question, it should be attributed to the effect of multiple interference between line-defects. When setting $m=5$ in Fig. 6(b), except $N_d=30$, the transmission coefficients of all other lines become active and manifest some peaks at some fixed incident angles. For the cases of $m=7$ in Fig. 6(c) and $m=10$ in Fig. 6(d), more transmission peaks appears and overlap each other. However, the electron transmission coefficient in case of $N_d=30$ decreases gradually with increasing the number of paralleling line-defects. It is because at $E=0.01$ electron's wave length is considerably larger than the distance of two nearest neighbouring line-defects, $N_d=30$.

For sake of concreteness, here we restrict to the case of six line-defects and make great efforts to manifest clearly a picture of a $100\%$ valley polarization. In Fig. 7, the transmission coefficients in two valleys are mapped in the panel of $(q_x,q_y)$. It is clearly shown that comparing to the case of two line-defects, more large transmission regions (dark red) exist in Fig. 7, the case of six line-defects. As explained in Fig. 4 and 5, these high transmission regions should likewise be from the resonant tunneling due to the quantum interference, which is also named as Fabry-P$\'{e}$rot interference in optics.

In Fig. 8, the transmission coefficient is plotted as a function of the incident angle $\alpha$, which follows half of the green path in Fig. 7. Note that here the electron momentum $q$ is taken as $q=0.015/a$. Obviously, the marking A, B, and C transmission peaks should be attributed to the three intersections between the green path and the large transmission regions (dark red) in Fig. 7. Here, $\alpha_{\small \small A}\approx0.38\pi$, $\alpha_{\small B}\approx0.15\pi$, and  $\alpha_C\approx -0.05\pi$ illustrate that the A transmission peak is attributed to the intrinsic scattering property of a line-defect and the peaks at B and C are from the resonant tunneling of the positive and negative first standing waves, respectively.

Much importantly, not only a perfectly resonant transmission but also a $100\%$ polarization can be obtained at the $B$ resonant peak by the polarization formula: $\mathcal{P}=\frac{T_{k}-T_{k'}}{T_{k}+T_{k'}}$. In a conclusion, we predict that the perfect $100\%$ polarization can be achieved by scattering off many line-defects, which origins in the quantum resonant tunneling phenomenon. Even though only exhibiting the cases of two and six line-defects here, in principle this high transmission and thus the $100\%$ polarization can be realized in graphene with arbitrary number of line-defects. This feature shed a light on constructing a feasible valley polarization resonator by using multiple line-defects.

\begin{figure}
\center
  \includegraphics[width=0.8\columnwidth]{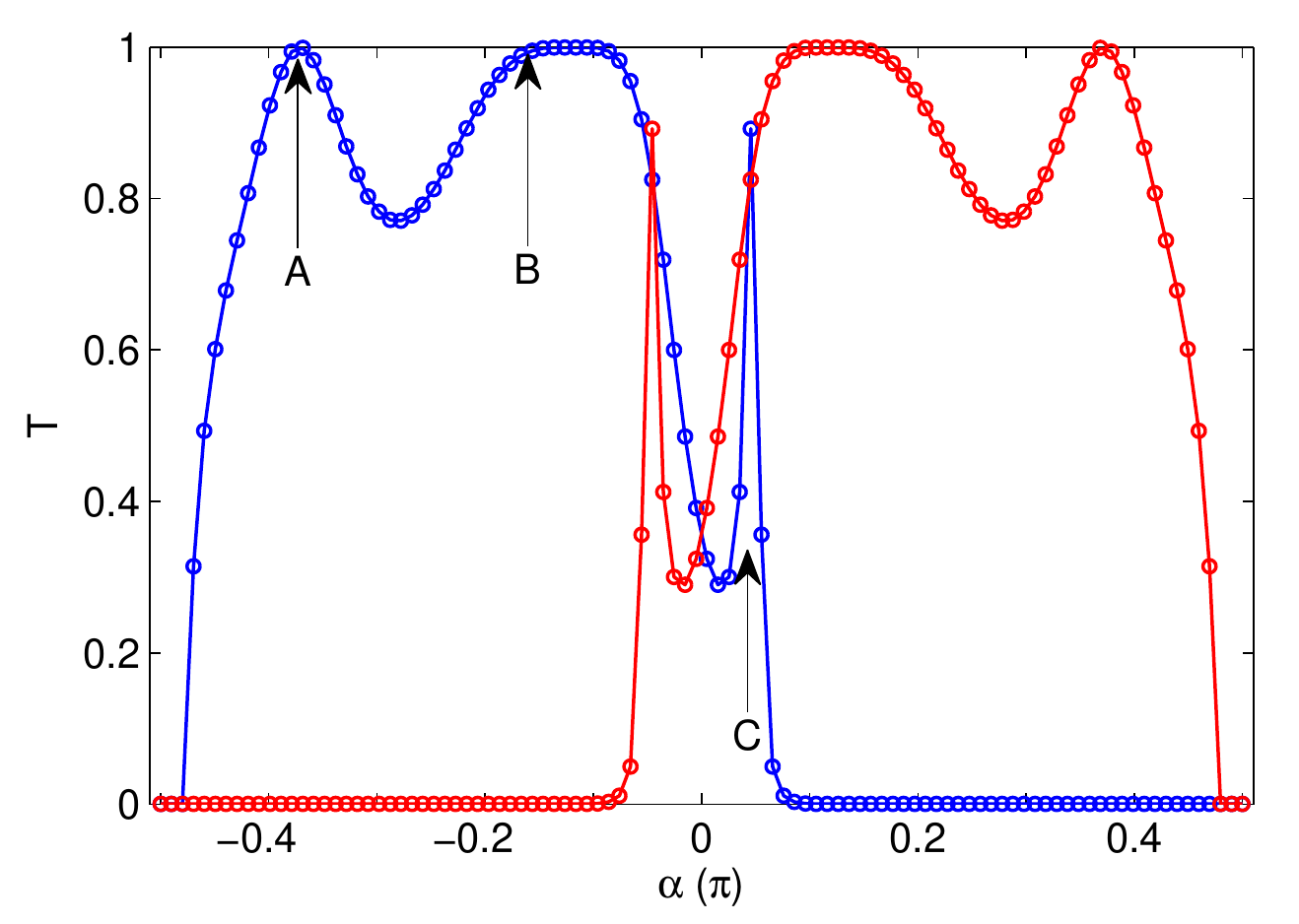}
\caption{ (Color online) The transmission coefficient with a fixed momentum $q=0.015/a$ which is exactly the green circle path in Fig. 6. Two lines (blue and red) correspond to that of $K$ and $K$ valleys, respectively. In this case, six line-defects are considered and the fixed distance between arbitrary two neighboring line-defects is $N_d=70$.}
\end{figure}
\section{Conclusions}
\label{sec:conclusions}
In summary, we study the valley polarization in the presence of several paralleling line-defects. Importantly, the valley polarization changes with varying the distance between the line-defects and the number of the line-defects. In particular, the valley polarization can reach to the maximum value of one, namely a 100$\%$ polarization, at a small incident angle. Further numerical calculations manifest that $100\%$ polarization should be attributed to the behavior of electron quantum interference, more precisely the phenomenon of the standing wave between the line-defects. The findings make us get rid of helplessness in the choose of the electron's incident angle and provides a reliable way to realize a controllable valley polarization in experiment. With the great advance in graphene related technologies, a valley polarization resonator and also a detector of a valley polarization, being the key components in valleytronics, are promising to be realized in the system of the line-defect graphene.

In the preparation of this paper, we notice a reference on two paralleling line-defect in recent\cite{Gunlycke2} and intriguingly, nearly a perfect perfect valley polarization is also observed. This report also illustrates that the valley filter of several paralleling line-defects should feasible in experiment.

\section*{ACKNOWLEDGMENTS}
Yang Liu and Juntao Song want to thank Liwei Jiang and Yisong Zheng for their help in the preparation of this paper. We gratefully acknowledge the financial support from NSF-China under Grant Nos. 11204065(J.T.S.), 11274089(Y.L.), 10974043(Y.X.L.), and 11274364(Q.F.S.). J.S. is also supported by RFDPHE-China under Grant No. 20101303120005.
\\

\section*{APPENDIX}
\renewcommand{\theequation}{A.\arabic{equation}}
In this appendix, we present a detailed derivation for the transmission coefficient $T$ of low energy electron in the vicinity of the Dirac point. Note that similar derivations can also reference to some previous papers.\cite{Zheng2,Zheng3}

In the Landauer-B\"{u}ttiker formula, Eq. (\ref{M4}), the Green's function $G^r$ can be readily expressed as: $G^r=1/(E-H_d-\Sigma^r_L-\Sigma^r_R)$. Here, $H_d$ is just a two-by-two matrix related to $\hat{T}_3$ and therefore the key point is to obtain the retarded self-energy function, $\Sigma_{L/R}^{r}$. That can be solved by the surface Green's function. In the following, we firstly present an analytic method of obtaining the surface Green's function $G^r_L$ and then derive the transmission coefficient of low energy electron in the vicinity of the Dirac point.

Based on the schematic representation in Fig. 1(c), a Hamiltonian of a semi-infinite right lead can be written as:
\begin{eqnarray}\label{A1}
\mathcal{H}=\left(
\begin{array}{cccccccc}
0                     &\mathcal{H}^{11}_{12}    &0                    &0                    & 0 & 0 & 0 & \cdots \\
\mathcal{H}^{11}_{21} &0                        &\mathcal{H}^{11}_{23}&0                    & 0 & 0 & \cdots \\
0                     &\mathcal{H}^{11}_{32}    &0                    &\mathcal{H}^{11}_{34}& 0 & 0 & 0 & \cdots \\
0                     &0                        &\mathcal{H}^{11}_{43}&0                    &\mathcal{H}^{12}_{41}& 0 & 0 & \cdots\\
0                     &0                        &0                    &\mathcal{H}^{21}_{14}&0                    &\mathcal{H}^{22}_{12}& 0                   &\cdots \\
0                     &0                        &0                    &0                    &\mathcal{H}^{22}_{21}&0                    &\mathcal{H}^{22}_{23}& \cdots \\
0                     &0                        &0                    &0                    & 0                   &\mathcal{H}^{22}_{32}&0                    & \cdots \\
\vdots                &\vdots                   &\vdots               &\vdots               & \vdots              &\vdots               &\vdots               & \ddots\\
\end{array}
\right),
\end{eqnarray}
where $H^{LL'}_{ll'}$ means a hopping term from the $l$ sub-layer in the $L$ supercell to the $l'$ sub-layer in the $L'$. Obviously, the relations hold true: $\mathcal{H}^{LL}_{12}=-\hat{T}_1$, $\mathcal{H}^{LL}_{23}=-\hat{T}_2$, $\mathcal{H}^{LL}_{34}=-\hat{T}^\dagger_1$, and $\mathcal{H}^{LL+1}_{41}=-\hat{T}^\dagger_2$.

Thus, the eigen equation $[\mathcal{H}][\mathcal{C}]=E[\mathcal{C}]$ can be written as:
\begin{eqnarray}\label{A2}
-E\mathcal{C}^1_1+\mathcal{H}^{11}_{12}\mathcal{C}^1_2=0,&&\nonumber\\
\mathcal{H}^{11}_{21}\mathcal{C}^1_1-E\mathcal{C}^1_2+\mathcal{H}^{11}_{23}\mathcal{C}^1_3=0,&&\nonumber\\
\mathcal{H}^{11}_{32}\mathcal{C}^1_2-E\mathcal{C}^1_3+\mathcal{H}^{11}_{34}\mathcal{C}^1_4=0,&&\nonumber\\
\mathcal{H}^{11}_{43}\mathcal{C}^1_3-E\mathcal{C}^1_4+\mathcal{H}^{12}_{41}\mathcal{C}^2_1=0,&&\\
\mathcal{H}^{21}_{14}\mathcal{C}^1_4-E\mathcal{C}^2_1+\mathcal{H}^{22}_{12}\mathcal{C}^2_2=0,&&\nonumber\\
\cdots\ \ \ \ \  && \nonumber
\end{eqnarray}
where $E$ and $[\mathcal{C}]$ represents the eigenvalue and eigenstate, respectively. For simplicity, we cast the above eigen equation into a subspace and consequently obtain the following expression,
\begin{eqnarray}\label{A3}
\mathcal{W}^{11}_{11}{\mathcal{C}}^1_1+\mathcal{W}^{11}_{14}\mathcal{C}^1_4=0,&&\nonumber\\
\mathcal{W}^{11}_{41}\mathcal{C}^1_1+\mathcal{W}^{11}_{44}\mathcal{C}^1_4+\mathcal{W}^{12}_{41}\mathcal{C}^2_1=0,&&\nonumber\\
\mathcal{W}^{21}_{14}\mathcal{C}^1_4+\mathcal{W}^{22}_{11}{\mathcal{C}}^2_1+\mathcal{W}^{22}_{14}\mathcal{C}^2_4=0,&&\\
\mathcal{W}^{22}_{41}\mathcal{C}^2_1+\mathcal{W}^{22}_{44}\mathcal{C}^2_4+\mathcal{W}^{23}_{41}\mathcal{C}^3_1=0,&&\nonumber\\
\cdots\ \ \ \ \  && \nonumber
\end{eqnarray}
Here,
\begin{eqnarray}\label{A4}
&&\mathcal{W}^{LL}_{11}=\mathcal{H}^{LL}_{11}+\mathcal{H}^{LL}_{12}E^{-1}\mathcal{H}^{LL}_{21}\nonumber\\
&&\ \ \ \ \ \ \ \ \ \ +\mathcal{H}^{LL}_{12}E^{-1}\mathcal{H}^{LL}_{23}\mathcal{F}^{-1}\mathcal{H}^{LL}_{32}E^{-1}\mathcal{H}^{LL}_{21},\nonumber\\
&&\mathcal{W}^{LL}_{14}=\mathcal{H}^{LL}_{12}E^{-1}\mathcal{H}^{LL}_{23}\mathcal{F}^{-1}\mathcal{H}^{LL}_{34},
\end{eqnarray}
where $\mathcal{F}^{-1}=E-\mathcal{H}^{LL}_{32}E^{-1}\mathcal{H}^{LL}_{23}$. Note that other quantities, such as $\mathcal{W}^{LL}_{44}$, $\mathcal{W}^{LL}_{41}$ and $\mathcal{W}^{LL+1}_{41}$, can be also obtained by using Dyson equation of Green's function.
For simplicity in the following derivation, we define some symbols as:
\begin{eqnarray}
&&\mathcal{\omega}_i=\mathcal{W}^{LL}_{11}=\mathcal{M}/(E^2-1),\nonumber\\
&&\mathcal{\omega}_o=\mathcal{W}^{LL}_{14}=E-\mathcal{M}E/(E^2-1),\\
&&\mathcal{\omega}_e=\mathcal{W}^{LL+1}_{41}=1,\nonumber
\end{eqnarray}
where $\mathcal{M}$ takes a form of
\begin{eqnarray}\label{A5}
\mathcal{M}=2\left(
\begin{array}{cc}
1                                &e^{i\theta}\cos{\theta}    \\
e^{-i\theta}\cos{\theta}         &1                        \\
\end{array}
\right).
\end{eqnarray}

Following the similar derivation processes of the eigen equation above, the Green's function can be written readily as:
\begin{eqnarray}\label{A6}
\omega_o{\mathcal{G}}^1_1+\omega_i\mathcal{G}^1_4=1,&&\nonumber\\
\omega_i\mathcal{G}^1_1+\omega_0\mathcal{G}^1_4+\omega_e\mathcal{G}^2_1=0,&&\nonumber\\
\omega_e\mathcal{G}^1_4+\omega_0{\mathcal{G}}^2_1+\omega_i\mathcal{G}^2_4=0,&&\\
\omega_i\mathcal{G}^2_1+\omega_0\mathcal{G}^2_4+\omega_e\mathcal{G}^3_1=0,&&\nonumber\\
\cdots\ \ \ \ \  && \nonumber
\end{eqnarray}
After a gauge transformation by a unitary matrix, which is denoted as:
\begin{eqnarray}\label{A7}
\mathcal{U}=\frac{1}{\sqrt{2}}\left(
\begin{array}{cc}
-e^{i\theta}   &e^{i\theta}    \\
1              &1              \\
\end{array}
\right),
\end{eqnarray}
the $\mathcal{M}$ matrix, all $\omega_{\delta}$ matrixes, and Green's function in Eq. (\ref{A6}) can be transformed into:
\begin{eqnarray}\label{A8}
&&\mathcal{U}^\dagger \mathcal{M} \mathcal{U}=\left(
\begin{array}{cc}
m_1   &0    \\
0     &m_2              \\
\end{array}
\right),\ \
\mathcal{U}^\dagger \omega_\delta \mathcal{U}=\left(
\begin{array}{cc}
\widetilde{\omega}_{\delta 1}   &0                  \\
0                   &\widetilde{\omega}_{\delta 2}  \\
\end{array}
\right),\nonumber \\
&&\mathcal{U}^\dagger \mathcal{G}^L_l \mathcal{U}=\left(
\begin{array}{cc}
\widetilde{g}^L_{l1} &0    \\
0  &\widetilde{g}^L_{l2}  \\
\end{array}
\right),
\end{eqnarray}
where $m_1=2(1-\cos\theta)$ and $m_2=2(1+\cos\theta)$.
Note that the Green's function must be diagonalized if all coefficients in Eq. (\ref{A6}), namely all $\omega_\delta$, can be diagonalized by this unitary transformation.

As such, Eq. (\ref{A6}) can be decomposed into:
\begin{eqnarray}\label{A9}
\left(
\begin{array}{cccccc}
\widetilde{\omega}_{o p}&\widetilde{\omega}_{i p}&0                       &0                       &0 &\cdots \\
\widetilde{\omega}_{i p}&\widetilde{\omega}_{o p}&\widetilde{\omega}_{e p}&0                       &0 &\cdots \\
0                       &\widetilde{\omega}_{e p}&\widetilde{\omega}_{o p}&\widetilde{\omega}_{i p}&0 &\cdots \\
0                     &0                         &\widetilde{\omega}_{i p}&\widetilde{\omega}_{o p}&\widetilde{\omega}_{e p}& \cdots\\
0                     &0                         &0                       &\widetilde{\omega}_{e p}&\widetilde{\omega}_{o p}&\cdots \\
\vdots                &\vdots                   &\vdots               &\vdots               & \vdots                 & \ddots\\
\end{array}
\right)
\left(
\begin{array}{c}
\widetilde{g}^1_{1p} \\
\widetilde{g}^1_{4p}\\
\widetilde{g}^2_{1p} \\
\widetilde{g}^2_{4p}\\
\widetilde{g}^3_{1p} \\
\vdots                \\
\end{array}
\right)=
\left(
\begin{array}{c}
1 \\
0 \\
0 \\
0\\
0\\
\vdots                \\
\end{array}
\right),
\end{eqnarray}
where $p$ takes $1$ or $2$.

There are many known mathematic method to calculate an inverse matrix of a tridiagonal matrix. However, we refrain from introducing these methods here.
We use $D_0$ to denote the determinant of the tridiagonal matrix on the left side of Eq. (\ref{A9}) and $D_i$ to denote the determinant of the tridiagonal submatrix except the former $i$ columns and rows of the tridiagonal matrix.\cite{Zheng3} Then, we can obtain the recursive equations:
$D_0/D_1=\widetilde{\omega}_{o}-\widetilde{\omega}^2_{i}D_2/D_1$ and $D_1/D_2=\widetilde{\omega}_{o}-\widetilde{\omega}^2_{e}D_3/D_2$. In the limit of infinite long lead, corresponding to an infinite large tridiagonal matrix in Eq. (\ref{A9}), the relation $D_0/D_1=D_2/D_3$ maintains. From the formula $\widetilde{g}^1_{1p}=D_1/D_0$, we can readily obtain
\begin{eqnarray}\label{10}
\widetilde{g}^1_{1p}=\frac{(\widetilde{\omega}^2_{op}+1-\widetilde{\omega}^2_{ip})
+\sqrt{(\widetilde{\omega}^2_{op}+1-\widetilde{\omega}^2_{ip})^2-4\widetilde{\omega}^2_{op}}}
{2\widetilde{\omega}_{op}},
\end{eqnarray}
where $\widetilde{\omega}_{ep}=1$ is adopted.

Performing a inverse unitary transformation with respect to Eq. (\ref{A8}), the surface Green's function of a semi-infinite long lead can be expressed as:
\begin{eqnarray}\label{11}
\mathcal{G}^1_1=
\mathcal{U}\left(
\begin{array}{cc}
\widetilde{g}^1_{11}   &0    \\
0                      &\widetilde{g}^1_{12}              \\
\end{array}
\right)\mathcal{U}^\dagger
=\frac{1}{2}\left(
\begin{array}{cc}
\mathcal{A}   &-\mathcal{B} e^{i\theta}    \\
-\mathcal{B} e^{-i\theta}       &\mathcal{A}              \\
\end{array}
\right),
\end{eqnarray}
where $\mathcal{A}=\widetilde{g}^1_{11}+\widetilde{g}^1_{12}$ and $\mathcal{B}=\widetilde{g}^1_{11}-\widetilde{g}^1_{12}$. Now, we get an exact expression about the surface Green's function of a semi-infinite long lead.

Combining the self energy function and the Green's function of the line-defect
\begin{eqnarray}\label{A12}
&&\Sigma^r_{L/R}=\tau^2_1\mathcal{G}^1_1\nonumber\\
&&G^r=1/(E+\tau_2\hat{T}_3-\Sigma^r_{L}-\Sigma^r_{R})
\end{eqnarray}
Substituting the above formula into the Landauer-B\"{u}ttiker formula, Eq. (\ref{M4}), we can express the transmission coefficient exactly as:
\begin{eqnarray}\label{A12}
&&T_{k_y}(E)=\frac{\tau^4_1 }{|F_2|^2}\big[(\textmd{Im}\widetilde{g}^1_{11})^2|F_1|^2+(\textmd{Im}\widetilde{g}^1_{12})^2|F_3|^2 \nonumber\\
&&\ \ \ \ \ \ \ \ \ \ \ +8\tau^2_2\textmd{Im}\widetilde{g}^1_{11}\textmd{Im}\widetilde{g}^1_{12}\cos^2\theta\big],
\end{eqnarray}
where $F_1=2E-4\tau^2_1\widetilde{g}^1_{12}+2\tau_2\cos{\theta}$, $F_2=(E-\mathcal{A}\tau^2_1)^2-\mathcal{B}\tau^2_1(\mathcal{B}\tau^2_1+2\tau_2\cos{\theta})-\tau^2_2$, and $F_3=2E-4\tau^2_1\widetilde{g}^1_{12}-2\tau_2\cos{\theta}$.

In the low energy limit, it is not difficult to find $\widetilde{g}^1_{11}\sim t$, $\widetilde{g}^1_{12}\sim E\sim 0$, and $\widetilde{g}^1_{12}\ll \widetilde{g}^1_{11}$. Thus, the transmission coefficient is approximately written as:\cite{Zheng2}
\begin{eqnarray}\label{A13}
&&T_{k_y}(E)\approx\frac{\tau^4_1 \textmd{Im}(\widetilde{g}^1_{11})^2|F_1|^2}{|F_2|^2}.
\end{eqnarray}
Note that we still reserve the term of $\textmd{Im}\widetilde{g}^1_{12}$ in $F_1$ and $F_2$ because some terms in $F_1$ and $F_2$ are in the same order as $\textmd{Im}\widetilde{g}^1_{12}$.

Assuming incident electron's energy is much smaller than the hopping terms $t$, $\tau_1$ and $\tau_2$, an approximate expression of the transmission coefficient can be written as:
\begin{eqnarray}\label{A14}
T_{k_y}(E)=\frac{\tau^4_1(1+\cos{2\theta})[4E^2-(1-m_1)^2]/2}{2\tau^2_1\tau_2\cos{\theta}(1-m_1)+E^2+4\tau^4_1\cos^2{\theta}E^2}
\end{eqnarray}
Following the relations: $k_x=q_x$, $k_y=\pm \pi/3a+q_y$, $E=\sqrt{3}qa/2$, $\theta=k_y a\approx\pm \pi/3$, and $\tan{\alpha}=q_y/q_x$, we adopt the approximation in the low energy limit: $1-m_1\approx\pm 2E\sin{\alpha}$, $1+\cos{2\theta}=1/2$, and $\cos{\theta}=1/2$ in Eq. (\ref{A14}). We can readily obtain
\begin{eqnarray}\label{A15}
T_{K/K'}(E)=\frac{1}{2}(1\pm\sin{\alpha}).
\end{eqnarray}
That is namely the transmission coefficient of low energy electron in the vicinity of two Dirac valleys.\cite{Gunlycke,Zheng2}
\
\\
\
\\

\end{document}